\documentclass[prl,onecolumn,superscriptaddress,amsmath,amssymb,showpacs,longbibliography]{revtex4-1}
\usepackage{hyperref}
\pdfoutput=1
\usepackage{graphicx}

\usepackage{amsmath}
\usepackage{bm}
\usepackage{amssymb}
\usepackage{textcomp}
\usepackage{stmaryrd}
\usepackage{ulem}
\usepackage{gensymb}

\usepackage{import}
\usepackage{color}
\usepackage{verbatim}

\usepackage{float}

\usepackage{sistyle} % Unité SI

\def\XXint#1#2#3{{\setbox0=\hbox{$#1{#2#3}{\int}$}
     \vcenter{\hbox{$#2#3$}}\kern-.5\wd0}}

\begin{document}

\title{Single- and narrow-line photoluminescence in a boron nitride-supported MoSe$_2$/graphene heterostructure}
%need to find a more compelling title...je cherche !

\author{Luis E. Parra L\'{o}pez}

\affiliation{Universit\'e de Strasbourg, CNRS, Institut de Physique et Chimie des Mat\'eriaux de Strasbourg (IPCMS), UMR 7504, F-67000 Strasbourg, France}

\author{Loïc Moczko}
\affiliation{Universit\'e de Strasbourg, CNRS, Institut de Physique et Chimie des Mat\'eriaux de Strasbourg (IPCMS), UMR 7504, F-67000 Strasbourg, France}

\author{Joanna Wolff}
\affiliation{Universit\'e de Strasbourg, CNRS, Institut de Physique et Chimie des Mat\'eriaux de Strasbourg (IPCMS), UMR 7504, F-67000 Strasbourg, France}

\author{Aditya Singh}
\affiliation{Universit\'e de Strasbourg, CNRS, Institut de Physique et Chimie des Mat\'eriaux de Strasbourg (IPCMS), UMR 7504, F-67000 Strasbourg, France}
\affiliation{Department of Physics, Indian Institute of Technology Delhi, 110016, New Delhi, India}

\author{Etienne Lorchat}
\affiliation{Universit\'e de Strasbourg, CNRS, Institut de Physique et Chimie des Mat\'eriaux de Strasbourg (IPCMS), UMR 7504, F-67000 Strasbourg, France}

\author{Michelangelo Romeo}
\affiliation{Universit\'e de Strasbourg, CNRS, Institut de Physique et Chimie des Mat\'eriaux de Strasbourg (IPCMS), UMR 7504, F-67000 Strasbourg, France}

\author{Takashi Taniguchi}
\affiliation{International Center for Materials Nanoarchitectonics,
National Institute for Materials Science,  1-1 Namiki, Tsukuba 305-0044, Japan}

\author{Kenji Watanabe}
\affiliation{Research Center for Functional Materials, National Institute for Materials Science, 1-1 Namiki, Tsukuba 305-0044, Japan}

\author{St\'ephane Berciaud}
\email{stephane.berciaud@ipcms.unistra.fr}
\affiliation{Universit\'e de Strasbourg, CNRS, Institut de Physique et Chimie des Mat\'eriaux de Strasbourg (IPCMS), UMR 7504, F-67000 Strasbourg, France}
\affiliation{Institut Universitaire de France, 1 rue Descartes, 75231 Paris cedex 05, France}

\begin{abstract} 

Heterostructures made from van der Waals (vdW) materials provide a template to investigate a wealth of proximity effects at atomically sharp two-dimensional (2D) heterointerfaces. In particular, near-field charge and energy transfer in vdW heterostructures made from semiconducting transition metal dichalcogenides (TMD) have recently attracted interest to design model 2D "donor-acceptor" systems and new optoelectronic components. Here, using of Raman scattering and photoluminescence spectroscopies, we report a comprehensive characterization of a  molybedenum diselenide (MoSe$_2$) monolayer deposited onto hexagonal boron nitride (hBN) and capped by mono- and bilayer graphene. Along with the atomically flat hBN susbstrate, a single graphene epilayer is sufficient to passivate the MoSe$_2$ layer and provides a homogeneous environment without the need for an extra capping layer. As a result, we do not observe photo-induced doping in our heterostructure and the MoSe$_2$ excitonic linewidth gets as narrow as 1.6~meV, approaching the homogeneous limit. The semi-metallic graphene layer neutralizes the 2D semiconductor and enables picosecond non-radiative energy transfer that quenches radiative recombination from long-lived states. Hence, emission from the neutral band edge exciton largely dominates the photoluminescence spectrum of the MoSe$_2$/graphene heterostructure. Since this exciton has a picosecond radiative lifetime at low temperature, comparable with the non-radiative transfer time, its low-temperature photoluminescence is only quenched by a factor of $3.3 \pm 1$ and $4.4 \pm 1$ in the presence of mono- and bilayer graphene, respectively. Finally, while our bare MoSe$_2$ on hBN exhibits negligible valley polarization at low temperature and under near-resonant excitation, we show that interfacing MoSe$_2$ with graphene yields a single-line emitter with degrees of valley polarization and coherence up to $\sim 15\,\%$.

\end{abstract}

\maketitle

\section{Introduction}

Semiconducting transition metal dichalcogenides (TMDs) are layered crystals that exhibit a unique set of optical and electronic properties. In particular, TMDs undergo an indirect-to-direct bandgap transition when thinned down to the monolayer limit \cite{Mak2010,Splendiani2010}. Enhanced Coulomb interactions combined with reduced dielectric screening in two dimensions endow mono and few-layer TMDs  with room-temperature stable excitonic manifolds~\cite{Wang2018,Goryca2019}. In addition, TMD monolayers possess a valley pseudo-spin degree of freedom that can be manipulated using schemes inspired by decades of developments in optically-controlled spin dynamics~\cite{Xu2014,Schaibley2016}. Moreover, the relatively easy isolation of TMD monolayers layers and their stacking with partner layered materials in van der Waals (vdW) heterostructures offer opportunities to tune their properties and discover new electronic and optical phenomena~\cite{Mak2016}, as well as new proximity effects~\cite{Ciorciaro2020,Lyons2020} at atomically-sharp heterointerfaces. These fundamental properties can potentially be harnessed for an emerging generation of optoelectronic~\cite{Mak2016,Massicotte2016} and valleytronic systems~\cite{Luo2017,Avsar2017}.

\par

A pivotal example are TMD/graphene vdW heterostructures. Indeed, combining the semi-metallic and optically transparent character of graphene~\cite{Castroneto2009,Mak2012} with the unique optical properties of TMDs~\cite{Wang2018} yields a platform for  light-emitting systems which, as opposed to bare TMD monolayers, retain a high degree of valley polarization and, importantly, of valley coherence up to room temperature~\cite{Lorchat2018}. The reason is that there is an efficient non-radiative transfer of photoexcited carriers and excitons from the TMD to graphene~\cite{He2014,Froehlicher2018,Yuan2018,Selig2019}. This coupling shortens the excitonic lifetime down to the picosecond range and may thus significantly quench photoluminescence (PL), in particular at room temperature~\cite{He2014,Yuan2018,Froehlicher2018}, where the effective excitonic lifetime of a bare TMD monolayer can exceed one nanosecond~\cite{Froehlicher2018}. An interesting situation arises at low temperatures, where the radiative lifetime of the neutral exciton~\cite{Robert2016,Fang2019} is of the same order as the exciton transfer time towards graphene~\cite{Lorchat2020}. In this case, neutral exciton emission is minimally quenched. In contrast, non-radiative transfer remains sufficiently fast to massively quench emission from all the other long-lived excitonic species (including charged excitons, spin-dark excitons, localized excitons, exciton-phonon replica,\dots~\cite{Wang2018}). Moreover, since the Dirac point of graphene lies within the bandgap of Mo- and W-based TMDs, the graphene layer acts as a broadband acceptor of electrons and holes for moderate doping levels below $\sim 10^{13}~\rm{cm^{-2}}$~\cite{Yu2009,Liang2013,Wilson2017}, resulting in a complete neutralization of the TMD \textit{in the dark} through static charge transfer~\cite{Hill2017}. The combination of these two filtering effects produces single-line PL spectra arising from the radiative recombination of neutral excitons~\cite{Lorchat2020}. 

Beyond these general considerations, the photophysics of 2D materials and a fortiori of TMD/graphene heterostructures subtly depends on extrinsic factors and it is therefore necessary to carefully control their local chemical and dielectric environment. For instance, investigations in high vacuum prevents physisorption of water and organic molecules. Besides, it is known that conventional transparent susbtrates such as SiO$_2$ can host trapped charges and favor photoinduced doping under an incident photon flux~\cite{Froehlicher2018,Ryu2010,Miller2015,Lin2019}. A workaround is to use a more inert material such as hexagonal boron nitride (hBN). Indeed, hBN has proven to be an invaluable vdW material ever since its introduction as a dielectric substrate to reveal the intrinsic electron transport properties of graphene~\cite{Dean2010}. Its atomically flat nature and optical transparency provides a smooth, homogeneous environment that preserves the intrinsic optical features of TMDs. A direct manifestation is the narrowing of the emission lines in hBN-capped samples~\cite{Cadiz2017,Ajayi2017}, with linewidths near $1\rm ~ meV$ that approach the homogeneous limit  at cryogenic temperatures~\cite{Back2018,Scuri2018,Fang2019,Zhou2020}.

In this paper, using a combination of PL and Raman spectroscopies, we investigate an hBN-supported MoSe$_2$ monolayer capped by a graphene flake containing  monolayer (1LG) and bilayer domains (2LG). We show that this minimal vdW assembly benefits both from  the capping properties of hBN and from the emission filtering effect of graphene, leading to the absence of photoinduced doping combined with single-line intrinsic excitonic emission with linewidths as low as 1.6 meV at 14~K. In addition, we measure finite degrees of valley polarization and, importantly, of valley coherence up to $\sim 15\,\%$ at cryogenic temperatures. These results are promising considering the particularly small valley contrasts reported in MoSe$_2$ samples~\cite{Wang2015b,Kioseoglou2016}.

%\section{Experimental details}

\section{Optical characterization at room temperature}

\begin{figure*}[th!] %!th
\begin{center}
\includegraphics[width=0.95\linewidth]{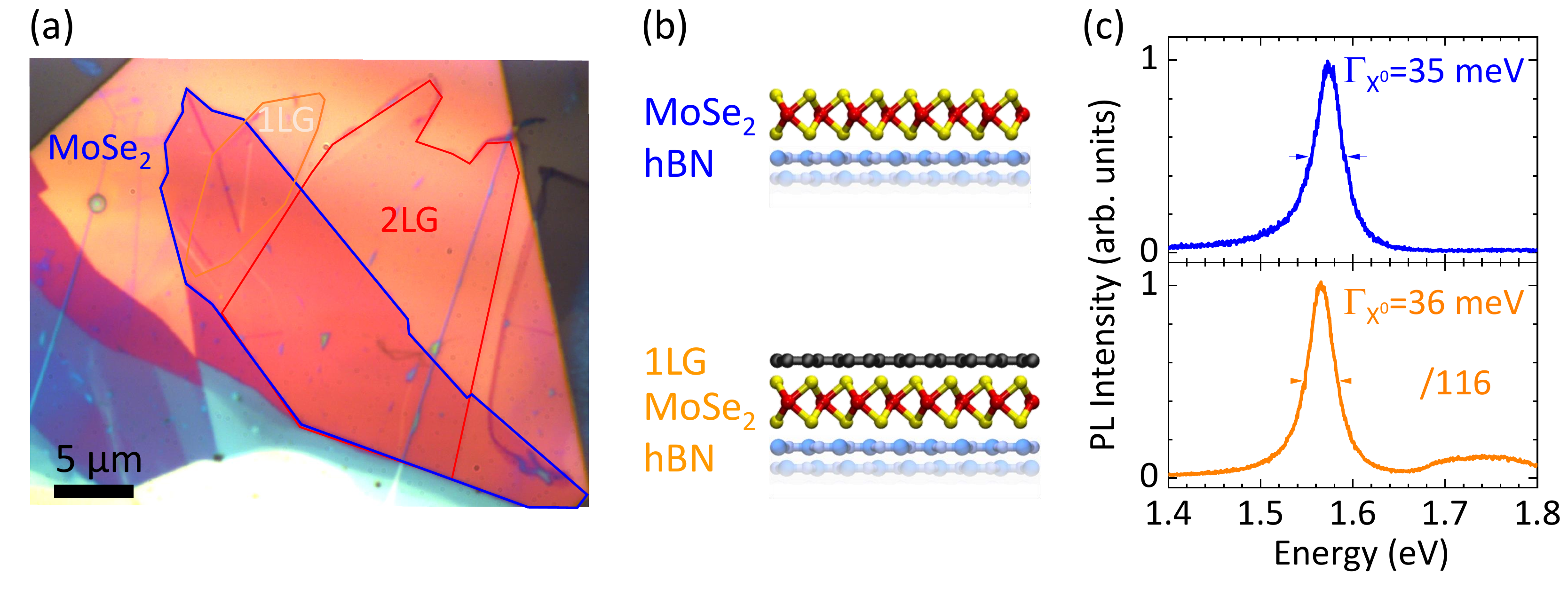}
\caption{\textbf{(a)} Optical image of a hBN-supported MoSe$_2$/graphene heterostructure deposited onto a Si/SiO$_2$ substrate. Three regions are contoured and correspond to hBN/MoSe$_2$ (blue), hBN/MoSe2$_2$/1LG (orange) and hBN/MoSe$_2$/2LG (red). \textbf{(b)} Sketch of the heterostructure showing the bare MoSe$_2$ (top) and the MoSe$_2$/1LG (bottom). \textbf{(c)} Room Temperature photoluminescence spectra recorded on bare MoSe$_2$ (top) and MoSe$_2$/1LG (bottom). The full-width at half maximum $\Gamma_{\rm X^0}$ of the main excitonic line is indicated. Both spectra are normalized to a common value and the scaling factor between the two spectra allowing an estimation of the room temperature PL quenching factor is indicated in the lower panel. The laser wavelength  is 532~nm and the laser intensity used in this experiment is near $1\mu\rm W/\mu m^{2}.$}
\label{Fig1}
\end{center}
\end{figure*}

%% Sample is normal + Quenching 

Figure \ref{Fig1}(a) shows an optical image of our hBN-supported MoSe$_2$/graphene sample  deposited onto a Si/SiO$_2$ substrate. The side view of the hBN/MoSe$_2$ and hBN/MoSe$_2$/1LG  are sketched in Fig.~\ref{Fig1}(b) and typical room temperature PL spectra these two systems are shown in Fig.~\ref{Fig1}(c). The main PL peak slightly below 1.6~eV arises from the recombination of the lowest lying optically active MoSe$_2$ exciton ($\rm X^0$, also known as the A-exciton) possibly with a minor contribution coming from the trion (charged exciton, $\rm X^{*}$)~\cite{Ross2013}. The photophysics of this heterostructure is driven by the efficient energy transfer from the thermalized excitonic population towards the graphene layer, leading to strong PL quenching by about two orders of magnitude (see scaling factor in Fig.~\ref{Fig1}(c)). As an indirect consequence of the picosecond lifetime of the band-edge exciton, we observe a broad, high energy shoulder on the MoSe$_2$/1LG PL spectrum that is assigned to hot luminescence from excited excitonic states, including Rydberg-like states from from the A-exciton manifold as well as the B-exciton (located near 190~meV above the band-edge A-exciton)~\cite{Froehlicher2018}. No sizeable hot PL signal is detected in the linear regime in the hBN-supported MoSe$_2$ reference.

\par

We now investigate the possibility of photoinduced charge transfer (i.e., photodoping) in the hBN/MoSe$_2$/1LG region of our sample. As indicated above, band offsets between TMD and graphene allow electron and hole transfer from the TMD to graphene. Besides this intrinsic phenomenon, the presence of surface traps and molecular adsorbates (either on the substrate or on the 2D material)~\cite{Ryu2010,Miller2015}, as well as chalcogen vacancies and other local defects in TMDs~\cite{Barja2019} can facilitate charge transfer and alter exciton dynamics. Under light illumination above the TMD bandgap, a net photoinduced charge transfer to graphene has been observed on SiO$_2$-supported samples on timescales that are orders of magnitude longer than the TMD excitonic lifetime, hence with no sizeable effect on the PL intensity~\cite{Froehlicher2018,Ahmed2020,Zhang2014}. This photodoping  leads to a stationary Fermi level shift in graphene and complementary fingerprints of charge transfer in the TMD monolayer~\cite{Froehlicher2018,Lin2019}. The Fermi level of graphene ultimately saturates as the photon flux increases~\cite{Froehlicher2018,Lin2019}. Although such extrinsic effects could  be of practical use for photodetection~\cite{Zhang2014,Ahmed2020}, they may hamper access to the intrinsic photophysics of the heterostructure. In our sample, the bottom hBN flake provides a flat substrate that passivates the MoSe$_2$/graphene heterostructure and could minimize extrinsic photodoping. Quantitative insights into photodoping can be obtained using micro-Raman spectroscopy~\cite{Ferrari2013} of the graphene flake. 

\par

\begin{figure*}[!ht]
\begin{center}
\includegraphics[width=1.0\linewidth]{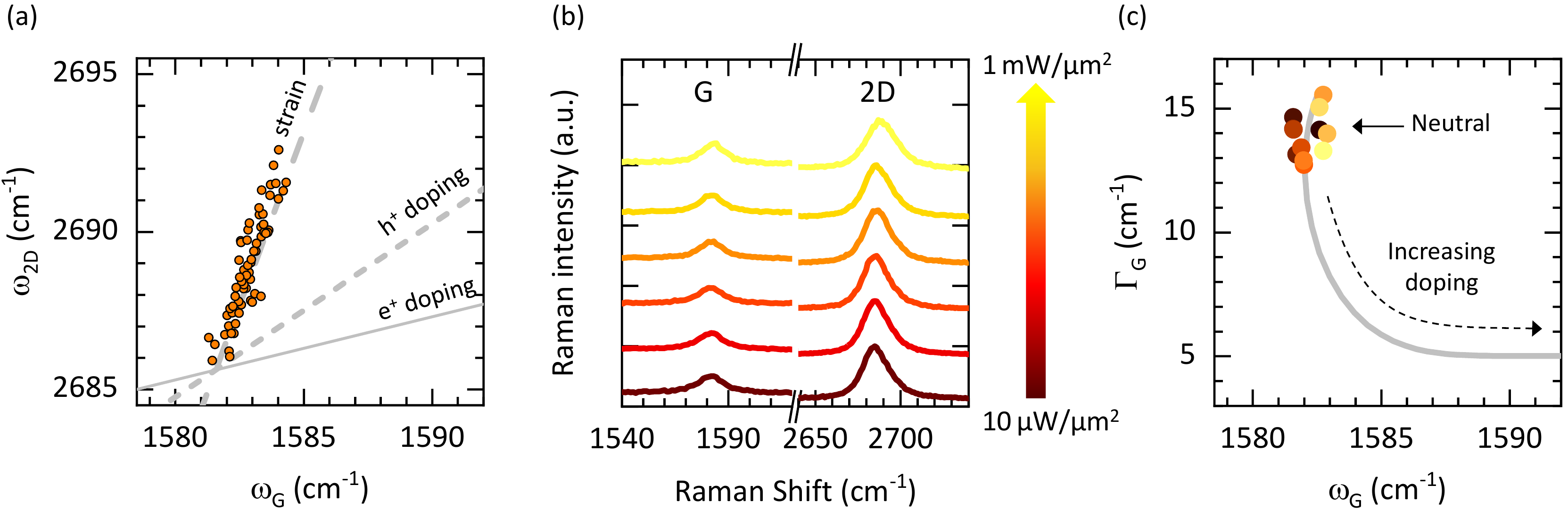}
\caption{(a) Correlation between $\omega_{\rm 2D}$ and $\omega_{\rm G}$ extracted from a Raman map of the region boxed in Fig.~\ref{Fig1}a. The dashed, short-dashed and solid gray lines indicate the correlations expected under biaxial strain, hole and electron doping, respectively. These lines cross at (1582,2686)~cm$^{-1}$, a point that corresponds to undoped graphene with a minimal amount of built-in compressive strain below $10^{-4}$.  Our measurements line up on the strain line, demonstrating the existence of compressive strain gradient on the graphene layer. (b) Typical Raman spectra taken with increasing, color coded, incident laser intensity. (c) Correlation between $\Gamma_{\rm G}$ and $\omega_{\rm G}$ under increasing incident laser intensity (color coded symbols, as in (b)). The solid gray line is the theoretically predicted correlation expected in the presence of doping using the model in Ref.~\cite{Pisana2007}. All measurements were performed in ambient conditions under laser excitation at 532~nm.}
\label{Fig2}
\end{center}
\end{figure*}

Indeed, the presence of charge carriers in 1LG leads to well-documented changes in the frequency ($\omega_{\rm G}$), full-width at half maximum (FWHM, $\Gamma_{\rm G}$) and intensity of its one-phonon G-mode (near $1582~\rm {cm^{-1}}$) due to the non-adiabatic renormalization of the Kohn anomaly at the $\mathbf{\Gamma}$ point. Under moderate doping below $10^{13}~\rm{cm^{-2}}$, this effect leads to an electron-hole symmetric  upshift of $\omega_{\rm G}$ accompanied by a reduction of $\Gamma_{\rm G}$ arising from the suppression of Landau damping~\cite{Pisana2007,Yan2007,Froehlicher2015a}. Similar, albeit less prominent effects are also observable in 2LG~\cite{Yan2008}. Contributions from doping can be disentangled from other perturbations, in particular due to built-in or applied strain, by inspecting the correlation between $\omega_{\rm G}$ and the frequency $\omega_{\rm 2D}$ of the 2D mode (near $2690~\rm {cm^{-1}}$, involving a pair of near-zone edge phonons with opposite momenta~\cite{Ferrari2013}). The slope $\frac{\partial{\omega_{\rm 2D}}}{\partial\omega_{\rm G}}$ is near 2.2 under biaxial strain, directly reflecting the values of the  Gr\"uneisen parameters 2D- and G-mode phonons~\cite{Lee2012,Metten2013,Metten2014} but is significantly smaller than 1 in the case of hole or electron doping (Fig.~\ref{Fig2}a) because the 2D-mode phonons have momenta significantly away from the Kohn anomaly at the zone edge (K and K' points)~\cite{Das2008,Froehlicher2015a}.

Figure \ref{Fig2}(a) shows the correlation between $\omega_{\rm G}$ and $\omega_{\rm 2D}$ extracted from a Raman map of the monolayer MoSe$_2$/1LG area of our sample (see Fig.~\ref{Fig1}a). Clearly, $\omega_{\rm 2D }$ and $\omega_{\rm G}$ follow a linear correlation with a slope near 2.2 (Fig.~\ref{Fig2}a), suggesting native compressive strain (i.e., stiffening of the Raman modes compared to the undoped/unstrained reference indicated in Fig.~\ref{Fig2}a) and ruling out sizable spatially inhomogeneous residual doping. The native strain level can be as large as $0.05~\%$ (if one assumes biaxial strain~\cite{Metten2014}) and the existence of a strain gradient likely arises from the stacking process. Figure \ref{Fig2}(b) shows the evolution of the Raman spectra as a function of the incident laser intensity from $10 ~ \rm{\mu W/\mu m^{2}}$ up to $1 ~\rm{mW /\mu m^{2}}$. We observe that $\omega_{\rm G}$ does not change appreciably and instead remains at $1582.2 \pm 0.5 ~ \rm{cm^{-1}}$. Similarly, $\Gamma_{\rm G}$ remains constant around $\approx 14 \pm 1~ \rm{cm^{-1}}$ as it can be seen in figure \ref{Fig2}(c). These values are typical for quasi-neutral graphene with a residual charge density of at most a few $10^{11}~\rm{cm^{-2}}$~\cite{Froehlicher2015a}. Hence, our observations strongly support the hypothesis that the hBN flake shields the heterostructure from substrate-induced electron redistribution and photo-doping, yielding a pristine system to investigate exciton dynamics.

\section{Low temperature hyperspectral PL mapping}

 \begin{figure*}[!h]
\begin{center}
\includegraphics[width=0.95\linewidth]{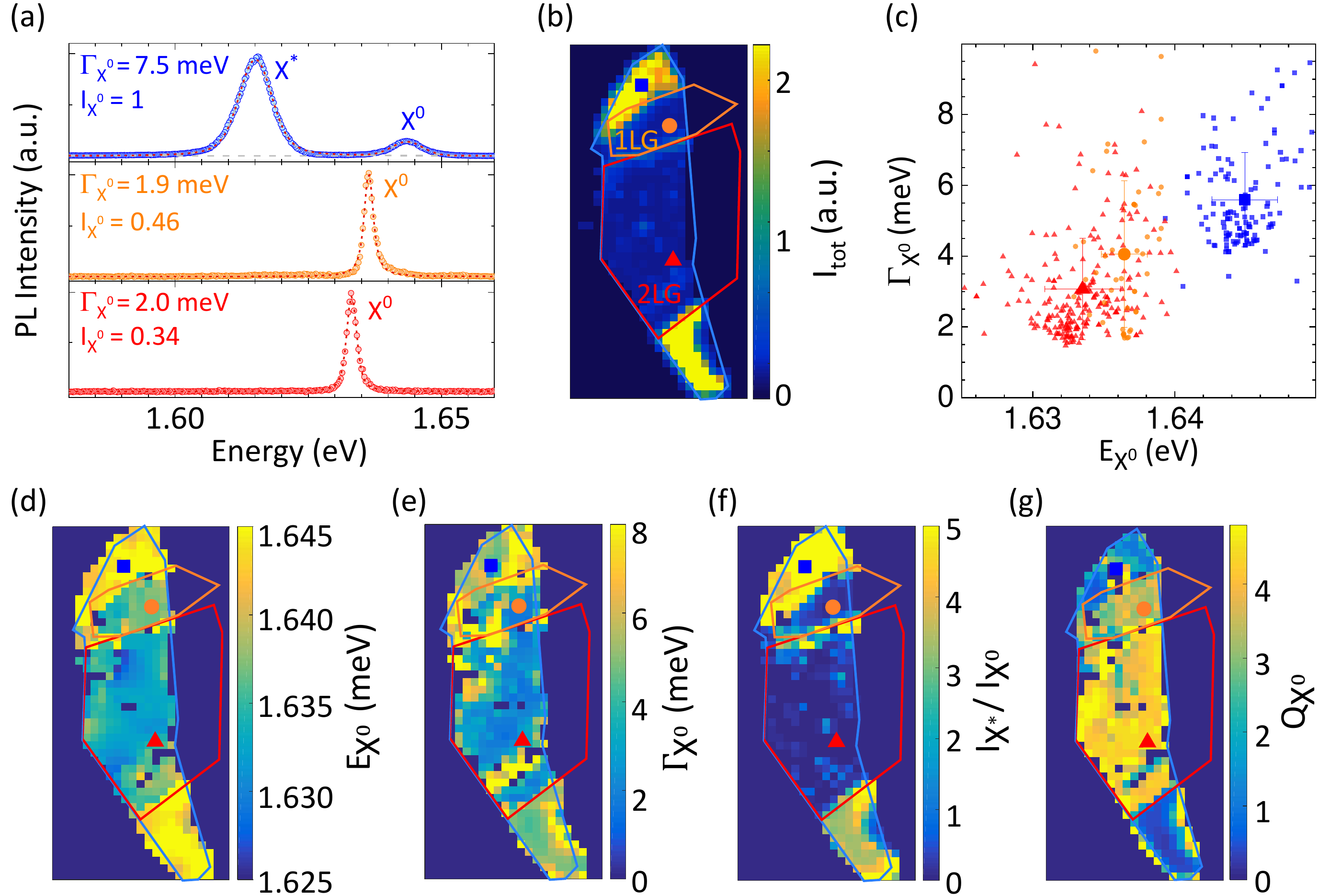}
\caption{Low-temperature photoluminescence of the MoSe$_2$/graphene van der Waals heterostructure shown in Fig.~\ref{Fig1}a. (a) From top to bottom, selected PL spectra taken in hBN-supported MoSe$_2$ (blue), MoSe$_2$/1LG (orange) and MoSe$_2$/2LG (red) at spots indicated in the PL intensity map shown in (b). Dashed red lines in (a) are Lorentzian fits to the data (filled circles). The integrated intensity of the $\rm X^0$ line is indicated in each case. Blue, orange and red contours in (b), (also in d-g) indicate the MoSe$_2$, 1LG and 2LG flakes, respectively. (c) Correlation between $\Gamma_{\rm X^0}$ and $E_{\rm X^0}$ extracted from the hyperspectral maps of the (d) neutral exciton emission energy $E_{\rm X^0}$, (e) full-width at half maximum $\Gamma_{\rm X^0}$, extracted from Voigt (bare MoSe$_2$) and Lorentzian (MoSe$_2$/1LG and MoSe$_2$/2LG) fits. The average values in each zone are shown  with their respective standard deviations following the color code in (a,b). (f) Map of the trion-to-exciton integrated intensity ratio $\frac{I_{\rm X^{*}}}{I_{\rm X^0}}$. (g) Map of the PL quenching factor of the $\rm X^0$ line. This factor, denoted $Q_{\rm X^0}$, is defined relative to the spatially averaged value of $I_{\rm X^0}$ over the MoSe$_2$ flake shown on the lower part of the map. The data in this figure were recorded at 14~K under continuous wave laser illumination at 633~nm with a laser intensity near 100~$\mathrm{\mu W/\mu m^2}$.}
\label{Fig3}
\end{center}
\end{figure*}

 Figure \ref{Fig3}a shows typical low temperature (14~K)  PL spectra recorded on selected spots of our sample indicated in the PL intensity map shown in Fig.~\ref{Fig3}b. An hyperspectral PL mapping study is shown in Fig.~\ref{Fig3}c-g. On the MoSe$_2$/hBN area, the PL spectrum is dominated by two main peaks located at $1.645\pm 0.002$ eV and  $1.616\pm 0.003$ eV (Fig.~\ref{Fig3}c,d). These peaks have Voigt profiles and originate from the recombination of the neutral exciton ($\rm X^0$) and the trion ($\rm X^{*}$), respectively~\cite{Ross2013}. In contrast, in the 1LG and 2LG-capped MoSe$_2$ regions, the PL spectra display single, narrower quasi-Lorentzian emission lines (Fig.~\ref{Fig3}c,e). As shown in Fig.~\ref{Fig3}a,c,d, these peaks are located at $1.636\pm 0.002$~eV for hBN/MoSe$_2$/1LG and $1.633\pm 0.003$~eV for hBN/MoSe$_2$/2LG, respectively. These values are slightly redshifted by 9 and 12 meV with respect to the bare MoSe$_2$ reference, respectively, due to dielectric screening~\cite{Raja2017,Froehlicher2018}.

 Considering the simplicity of the PL-spectrum of bright TMDs such as MoSe$_2$~\cite{Ross2013,Wang2018}, the single-line character of the MoSe$_2$/graphene PL spectra is essentially a consequence of the complete neutralization of the MoSe$_2$ monolayer. As confirmed by Fig.~\ref{Fig3}a,f, the integrated PL intensity near the expected location of the $\rm X^{*}$ feature is vanishingly small compared to that of $\rm X^0$  line ($I_{\rm X^0}$) over all the 1LG and 2LG-capped MoSe$_2$ area.

 The observed PL quenching, however, is a consequence of non-radiative energy transfer to graphene, stemming predominantly from hot excitons (either finite momentum $\rm X^0$ excitons residing outside the light cone or excitons occupying higher-order optically active states), with a smaller contribution from cold $\rm X^0$ excitons in the light cone~\cite{Lorchat2020}. As a result, the selected PL spectra in Fig.~\ref{Fig3}a and total PL intensity map ($I_{\rm {tot}}$, Fig.~\ref{Fig3}b) reveal sizeable reduction of $I_{\rm {tot}}$ on the MoSe$_2$/1LG and MoSe$_2$/2LG regions. However, if we consider the quenching factor of the neutral exciton line $\rm X^0$ only (denoted $Q_{\rm X^0}$, see Fig.~\ref{Fig3}g), we observe a moderate $Q_{\rm X^0}=3.3\pm 1$  on  MoSe$_2$/1LG and a slightly larger value of $4.4\pm 1$ in the MoSe$_2$/2LG, in qualitative agreement with previous studies of energy transfer from individual emitters to mono and few-layer graphene~\cite{Chen2010}.   The reduced quenching efficiency compared to the massive quenching observed at room temperature stems from the picosecond $\rm X^{0}$ radiative lifetime at low temperatures~\cite{Robert2016,Fang2019}, which is comparable to the energy transfer time to graphene (see Ref.~\cite{Lorchat2020} for details).

Our spatially-resolved PL study also reveals narrow $\rm X^0$ lines with spatially averaged FWHM ($\Gamma_{\rm X^0}$) of $4\pm 2~\rm{meV}$ and $3 \pm 1 ~\rm{meV}$, on the MoSe$_2$/1LG and MoSe$_2$/2LG areas, respectively (Fig.~\ref{Fig3}c,e). Interestingly, $\Gamma_{\rm X^0}$ gets as narrow as $1.6~\rm{meV}$ in selected locations (see Fig.~\ref{Fig3}e), a value that, even if it exceeds the homogeneous limit ($0.3~\rm{meV}$ for an estimated $\rm X^0$lifetime of 2~ps) suggests that our simple sample architecture warrants reduced inhomogeneous broadening and dephasing, with values that are close to state of the art hBN-encapsulated MoSe$_2$ samples~\cite{Fang2019,Zhou2020}. Figure \ref{Fig3}(c) presents the correlation between $\Gamma_{\rm X^0}$ and $E_{\rm X^0}$ in each region of the sample, extracted from the maps in Fig.~\ref{Fig3}d,e. As expected, the presence of a second graphene layer improves capping and further screens the MoSe$_2$ excitons.

 \begin{figure*}[!h]
\begin{center}
\includegraphics[width=0.55\linewidth]{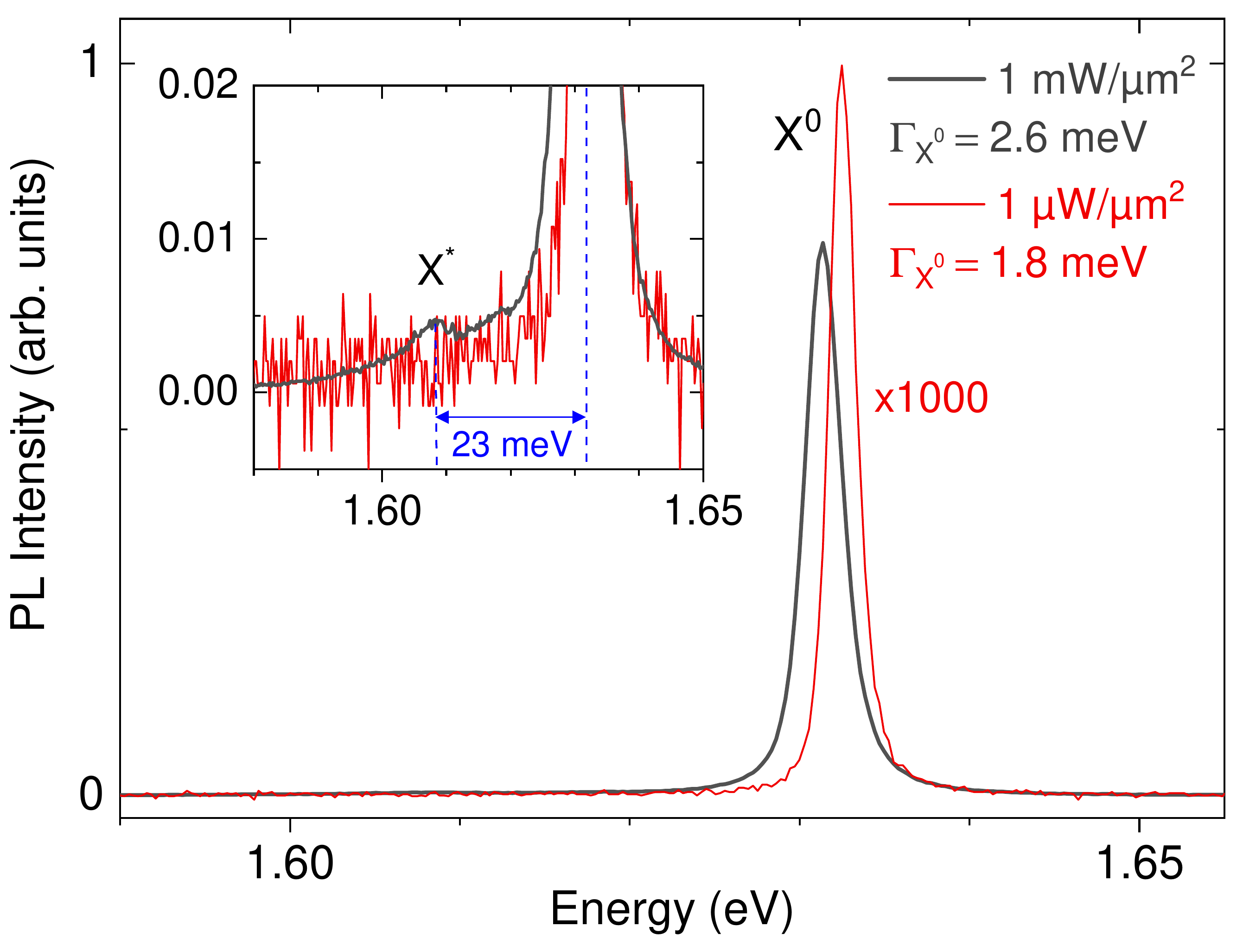}
\caption{Photoluminescence spectra of hBN/MoSe$_2$/2LG recorded under low  and high laser intensity, near $1~\rm{\mu W/\mu m^2}$ and $1~\rm{mW/\mu m^2}$ in red and grey, respectively. Both spectra were recorded under cw excitation at 633~nm with the same integration time. The low intensity spectrum has been multiplied by a factor of 1000 for comparison.  A fit of the data reveals linewidths (FWHM) of $\Gamma_{\rm X^0}=1.8$ and $2.6~\rm {meV}$ at low and high intensity, respectively and an integrated intensity ratio of the exciton line very near 1000, warranting that the sample remains in the linear regime under intense cw excitation.  The inset shows the low energy side of the spectra and evidences a faint feature assigned to a photoinduced trion $\rm X^{*}$ (see ref.~\cite{Lorchat2020} for details). The data were recorded under cw excitation at 633 nm.}
\label{Fig3bis}
\end{center}
\end{figure*}

Let us note that although our sample is spatially homogeneous over domains as large as 50~$\mu \rm m^2$, we observe more scattered $E_{\rm X^0}$, larger $\Gamma_{\rm X^0}$ and residual trion emission (Fig.~\ref{Fig3}d-g) near the boundaries of the 1LG and 2LG domains and also near localized micro-sized spots of our sample. During the stacking process, organic residues tend to segregate and form bubbles which locally decouple the TMD and graphene layers leading to complex spectra that reflect spatial inhomogeneities.

To close this section, let us comment on photoinduced doping at low-temperature. As indicated in Ref.~\cite{Lorchat2020}, TMD/graphene heterostructures are extremely photostable, in stark contrast with bare TMD monolayers. By illuminating our hBN/MoSe$_2$/2LG sample at high laser intensity ($\sim 1~\mathrm{mW/\mu m}^2$), a dim photo-induced trion feature with a reduced binding energy of 23~meV emerges, with an integrated intensity nearly three orders of magnitude smaller than that  of the $\rm X^{0}$ line (Fig.~\ref{Fig3bis}). These results indicate that the MoSe$_2$ layer coupled to graphene remains essentially neutral under intense cw illumination, and that one can safely neglect photo-induced doping, including at cryogenic temperatures.

\section{Valley polarization and valley coherence}

\begin{figure}[!ht]
\begin{center}
\includegraphics[width=0.9\linewidth]{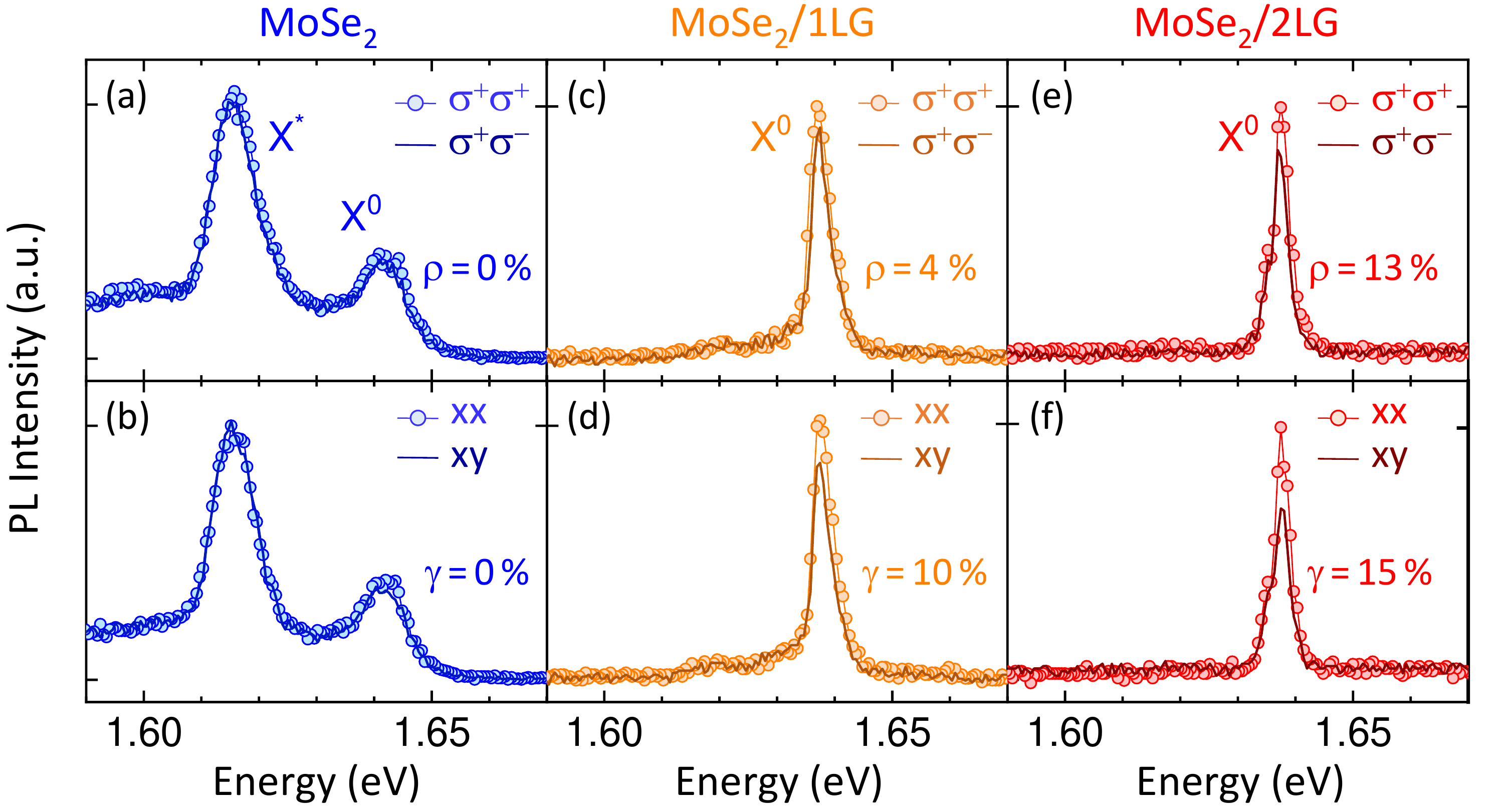}
\caption{Low-temperature polarization-resolved photoluminescence spectra in hBN-supported MoSe$_2$ (a,b), and MoSe$_2$/1LG (c,d) and MoSe$_2$/2LG (e,f). Measurements were taken at 9 K and under optical excitation at 60~meV above the $\rm X^0$ line with both circularly ($\sigma^\pm$) and linearly (x, y,) polarized laser excitation. Spectra in solid lines (resp. thin solid lines + symbols) correspond to cross-polarized (resp. co-polarized) incoming and emitted photons. The associated degrees of linear polarization ($\rho$, corresponding to the degree of valley polarization, in a,c,e) and circular polarization ($\gamma$, corresponding to the degree of valley coherence, in b,d,f) are indicated. The laser intensity is near 50~$\mathrm{\mu W/\mu m^2}$ and we have verified that all regions of the sample were excited in the linear regime. The spectra are normalized for clarity and the quenching factors of the $\rm X^{0}$ line are $Q_{\rm X^0}=1.8\pm0.1$ and $2.0\pm0.1$, respectively.}
\label{Fig4}
\end{center}
\end{figure}

We finally address the valley contrasting properties of our sample. TMD/graphene heterostructures have recently been introduced as chiral emitters with large degrees of valley polarization ($\rho$) and coherence ($\gamma$) up to room-temperature. Indeed, picosecond energy transfer between TMD and graphene allows the excitonic population to recombine before undergoing intervalley scattering and dephasing processes~\cite{Lorchat2018,Paradisanos2020}. In addition, graphene, as a capping material, reduces disorder, spatial inhomogeneities and dephasing, which explains the large room temperature value of $\gamma\approx 20\%$ recently reported in hBN-encapsulated WS$_2$/1LG heterostructures~\cite{Lorchat2018}. Although improved valley contrasts come at the expense of strong PL quenching at room temperature, we expect, in keeping with the discussion above, that sizeable values of $\rho$ and $\gamma$ combined with large $\rm X^{0}$ emission yields can be achieved at lower temperatures.

Fig.~\ref{Fig4} shows polarization-resolved PL spectra recorded using a circular and a linear basis, as well as the associated degrees of circular and linear polarization, which are directly yielding $\rho$ and $\gamma$, respectively. For the $\rm X^0$ emission line, we get $\rho\approx 0~\%$, $4\pm 4~\%$ and $13\pm 3~\%$  and $\gamma\approx 0~\%$, $10\pm 5~\%$ and $15\pm 4~\%$ for hBN-supported MoSe$_2$, MoSe$_2$/1LG and MoSe$_2$/2LG, respectively.

Near-zero values of $\rho$ have been reported in MoSe$_2$ monolayers~\cite{Wang2015b}.  Since the low temperature $\rm X^{0}$ lifetimes in bare MoSe$_2$ and in MoSe$_2$/graphene heterostructures  are similar ~\cite{Fang2019,Lorchat2020}, the observation of modest values of $\rho$ in MoSe$_2$/graphene is not surprising. Noteworthy, $\gamma$ significantly exceeds $\rho$ in MoSe$_2$/1LG and MoSe$_2$/2LG, suggesting significantly reduced dephasing in encapsulated samples, as recently observed in MoS$_2$ monolayers fully encapsulated in hBN~\cite{Cadiz2017} and in hBN-capped WS$_2$/1LG~\cite{Lorchat2018}. 
Here, our measurements are performed at 60~meV above the $\rm X^{0}$ line. We anticipate larger values of $\rho$ and $\gamma$ as we excite MoSe$_2$/graphene samples closer to the $\rm X^0$ resonance. In addition, one may also optimize the sample geometry as in Ref.~\cite{Fang2019,Zhou2020} in order to minimize the radiative lifetime of the MoSe$_2$ monolayer, and hence optimize $\rho$ and $\gamma$.

\par

%% Currently working on this
%%-----------------------------------
%%Not definitive version

\section{Conclusion and perspectives}   
We have demonstrated that hBN-supported TMD/graphene heterostructures offer an excellent template to study electron redistribution and excitonic energy transfer in two dimensions, with donor-acceptor distances in the sub-nanometer range. The absence of photodoping in hBN-supported samples as compared to SiO$_2$-supported systems adds on to the merits of hBN as an ideal dielectrics for van der Waals assembly. Our study also establishes that bright TMD emission with linewidths approaching the homogeneous limit can be achieved over large areas using only one epilayer of graphene, hence without the need of an hBN capping layer. In this manner, one does not only benefit from the emission filtering properties of graphene but also from its assets as a capping van der Waals material.

In spite of recent advances~\cite{Froehlicher2018,Lorchat2020,Yuan2018,Aeschlimann2020,Selig2019} the microscopic details of the transfer mechanism are still under investigation. Focusing for instance on energy transfer, Dexter- and F\"orster-type mechanisms depend sensitively on the distance between the layers and on the local environment. Investigations of the transfer efficiency  while finely tuning the TMD-graphene distance using hBN spacers and as a function of the number of graphene layers should bring decisive insights into near-field coupling in van der Waals materials.%    A distance-dependence study of the coupling between TMD and graphene can be useful to conclude on the mechanism behind the energy transfer. To do this, we can use hBN as a spacer to precisely control the separation.%%Not convincing enough.. need to work on this  

Our work also holds promise for device-oriented research. Indeed, the absence of a top hBN layer allows us to directly contact the graphene electrode for photodetectors and light-emitting devices, as well as for local investigations, e.g., using scanning tunnelling microscopy~\cite{Pommier2019}. Conversely, TMDs have recently been shown to outperform hBN as a capping material, yielding record-high room temperature electron mobility in graphene~\cite{Banszerus2019}.  

Finally, a challenging yet appealing perspective would consist in exploiting low-energy surface plasmon polaritons in graphene to tailor light-matter interactions in near-field coupled two-dimensional semiconductors~\cite{Kurman2018}, including in quantum wells made from the latter~\cite{Schmidt2018}.

\section{Experimental details}
Our van der Waals heterstructure was prepared using standard micromechanical exfoliation and dry transfer methods, as in ref~\cite{Castellanos2014}.
The hBN flake was mechanically exfoliated on top of the substrate. A MoSe$_2$ monolayer and a single- and bilayer graphene flake were mechanically exfoliated from bulk crystals and then stacked on top of the hBN flake. The number of layers of the flakes were verified using room-temperature Raman and PL spectroscopy. 
PL and Raman scattering measurements in Figs.~\ref{Fig1} and~\ref{Fig2}, respectively, were performed in ambient air using a 532~nm diode pumped solid state laser and a home-built scanning confocal setup. Low temperature PL studies were performed in a continuous-flow liquid helium cryostat mounted onto our Raman/PL microscope. All PL measurements in Fig.~\ref{Fig3} were recorded in the linear regime under cw excitation at $633~\rm {nm}$, i.e., slightly above the B exciton in MoSe$_2$~\cite{Froehlicher2018,Wang2018}. The low-temperature polarization-resolved measurements in Fig.~\ref{Fig4} were performed under near-resonant optical excitation 60~meV above the neutral exciton ($\rm X^0$) line using a tunable supercontinuum laser. The degree of valley polarization $\rho=\frac{I_{\sigma^+\sigma^+}-I_{\sigma^+\sigma^-}}{I_{\sigma^+\sigma^+}+I_{\sigma^+\sigma^-}}$ is measured  in a circular basis and the degree of valley coherence $\gamma=\frac{I_{xx}-I_{xy}}{I_{xx}+I_{xy}}$ is measured in a linear basis. In these expressions, $I_{ij}$ indicates the PL intensity detected in the "i" polarization state when upon excitation under  "j" polarized light (i,j= $\sigma^\pm$ or x/y). A Linear polarizer and an achromatic quarter-waveplate  were used to prepare the polarization state of the incoming beam. The emitted photons propagate through the same quarter-wave plate and are analyzed using a Wollaston prism, which permits to record simultaneously the co- and cross polarized PL signals.  The values of $\rho$ and $\gamma$ were determined from fits of the $\rm X^{0}$ lines.

\section*{Acknowledgements}
The authors thank X. Marie, C. Robert, D. Lagarde, S. Azzini, T. Chervy, C. Genet and G. Schull for fruitful discussions. We are grateful to Aditi Moghe and to the StNano clean room staff for technical support. We acknowledge financial support from the Agence Nationale de la Recherche (under grants 2D-POEM ANR-18-ERC1-0009) and from the LabEx NIE (ANR-11-LABX-0058-NIE, under grant SPE2D). A.S. and S.B. acknowledge support for the Indo-French Centre for the Promotion of Advanced Research (CEFIPRA). K.W. and T.T. acknowledge support from the Elemental Strategy Initiative conducted by the MEXT, Japan ,Grant Number JPMXP0112101001,  JSPS
KAKENHI Grant Number JP20H00354 and the CREST(JPMJCR15F3), JST.

% The next command determines the bibliography style. Please do not
% change this.
%\bibliographystyle{crplain}
%\bibliographystyle{crunsrt}

%This calls all references from the .bib

%  This inserts the bib file

~

%\bibliography{Manualbib}

\end{document}